\documentclass[conference,onecolumn]{IEEEtran}
\IEEEoverridecommandlockouts
\usepackage[final]{graphicx}
\usepackage{cite}
\usepackage{amssymb}
\usepackage{amsmath}
\usepackage{amsthm}
\usepackage{amsfonts}
\usepackage{bm}
\usepackage{filecontents}
\usepackage{subfigure}
\usepackage{subfigmat}
\usepackage{makecell}
\usepackage{booktabs}
\usepackage{accents}
\usepackage{algorithm}
\usepackage{algpseudocode}
\usepackage{algorithmicx}
\usepackage[bottom=1 in,left=.625 in, right=.625 in,top=.7 in]{geometry}

\begin{document}
\title{\huge{An Approximate Solution for Symbol-Level Multiuser Precoding Using Support Recovery}}

\author{Alireza~Haqiqatnejad,~Farbod~Kayhan,~and Bj\"{o}rn~Ottersten\\
	Interdisciplinary Centre for Security, Reliability and Trust (SnT),~University of Luxembourg \\
	email: \{alireza.haqiqatnejad,farbod.kayhan,bjorn.ottersten\}@uni.lu
		\thanks{\fontsize{7.5}{8.5}\selectfont{The authors are supported by the Luxembourg National Research Fund (FNR) under CORE Junior project: C16/IS/11332341 Enhanced Signal Space opTImization for satellite comMunication Systems (ESSTIMS).}}}

\newtheorem{theorem}{Theorem}
\newtheorem{acknowledgement}[theorem]{Acknowledgement}
\newtheorem{axiom}[theorem]{Axiom}
\newtheorem{case}[theorem]{Case}
\newtheorem{claim}[theorem]{Claim}
\newtheorem{conclusion}[theorem]{Conclusion}
\newtheorem{condition}[theorem]{Condition}
\newtheorem{conjecture}[theorem]{Conjecture}
\newtheorem{corollary}[theorem]{Corollary}
\newtheorem{criterion}[theorem]{Criterion}
\newtheorem{definition}[theorem]{Definition}
\newtheorem{example}[theorem]{Example}
\newtheorem{exercise}[theorem]{Exercise}
\newtheorem{lemma}[theorem]{Lemma}
\newtheorem{notation}[theorem]{Notation}
\newtheorem{problem}[theorem]{Problem}
\newtheorem{proposition}[theorem]{Proposition}
\newtheorem{remark}[theorem]{Remark}
\newtheorem{solution}[theorem]{Solution}
\newtheorem{summary}[theorem]{Summary}
\newtheorem{property}{Property}

\newcommand{\defeq} {\overset{\underset{\mathrm{def}}{}}{=}}
\newcommand{\Deee} {\mathrm{\pmb{\delta}}}
\newcommand{\lamb} {\mathrm{\pmb{\Lambda}}}
\newcommand{\psiii} {\mathrm{\pmb{\psi}}}
\newcommand{\ups} {\mathrm{\pmb{\upsilon}}}
\newcommand{\Tht} {\mathrm{\pmb{\Theta}}}
\newcommand{\Theee} {\mathrm{\pmb{\theta}}}
\newcommand{\Lamb} {\mathrm{\pmb{\Lambda}}}
\newcommand{\g} {\mathrm{\pmb{\gamma}}}
\newcommand{\Gammaaa}{\mathrm{\pmb{\Gamma}}}
\newcommand{\Del}{\mathrm{\pmb{\Lambda}}}
\newcommand{\Sigmaaa}{\mathrm{\pmb{\Sigma}}}
\newcommand{\Siii}{\mathrm{\pmb{\psi}}}
\newcommand{\Varrr}{\mathrm{\pmb{\vartheta}}}
\newcommand{\Phiii}{\mathrm{\pmb{\phi}}}
\newcommand{\chiii}{\mathrm{\pmb{\chi}}}
\newcommand{\Mu} {\mathrm{\pmb{\mu}}}
\newcommand{\omg} {\mathrm{\pmb{\omega}}}
\newcommand{\HHH} {\mathrm{\pmb{H}}}
\newcommand{\QQQ} {\mathrm{\pmb{Q}}}
\newcommand{\bbb}{\mathrm{\pmb{b}}}
\newcommand{\uuu}{\mathrm{\pmb{u}}}
\newcommand{\ddd}{\mathrm{\pmb{d}}}
\newcommand{\gggg}{\mathrm{\pmb{g}}}
\newcommand{\EEE}{\mathrm{\pmb{E}}}
\newcommand{\WWW}{\mathrm{\pmb{W}}}
\newcommand{\BBB}{\mathrm{\pmb{B}}}
\newcommand{\ZZZ}{\mathrm{\pmb{Z}}}
\newcommand{\I}{\mathrm{\pmb{I}}}
\newcommand{\J}{\mathrm{\pmb{J}}}
\newcommand{\A}{\mathrm{\pmb{A}}}
\newcommand{\DDD}{\mathrm{\pmb{\Sigma}}}
\newcommand{\FFF}{\mathrm{\pmb{F}}}
\newcommand{\G}{\mathrm{\pmb{G}}}
\newcommand{\T}{\mathrm{T}}
\newcommand{\F}{\mathrm{F}}
\newcommand{\HH}{\mathrm{H}}
\newcommand{\REAL}{\mathrm{Re}}
\newcommand{\IMAG}{\mathrm{Im}}
\newcommand{\LLL}{\mathrm{\pmb{L}}}
\newcommand{\R}{\mathrm{\pmb{R}}}
\newcommand{\YYY}{\mathrm{\pmb{Y}}}
\newcommand{\CCC}{\mathrm{\pmb{C}}}
\newcommand{\XXX}{\mathrm{\pmb{X}}}
\newcommand{\MMM}{\mathrm{\pmb{M}}}
\newcommand{\PPP}{\mathrm{\pmb{P}}}
\newcommand{\RRR}{\mathrm{\pmb{R}}}
\newcommand{\GGG}{\mathrm{\pmb{G}}}
\newcommand{\TTT}{\mathrm{\pmb{T}}}
\newcommand{\OOO}{\mathrm{\pmb{0}}}
\newcommand{\aaa}{\mathrm{\pmb{a}}}
\newcommand{\h}{\mathrm{\pmb{h}}}
\newcommand{\qqq}{\mathrm{\pmb{q}}}
\newcommand{\eee}{\mathrm{\pmb{e}}}
\newcommand{\s}{\mathrm{\pmb{s}}}
\newcommand{\vvv}{\mathrm{\pmb{v}}}
\newcommand{\fff}{\mathrm{\pmb{f}}}
\newcommand{\zzz}{\mathrm{\pmb{z}}}
\newcommand{\ccc}{\mathrm{\pmb{c}}}
\newcommand{\x}{\mathrm{\pmb{x}}}
\newcommand{\yyy}{\mathrm{\pmb{y}}}
\newcommand{\ppp}{\mathrm{\pmb{p}}}
\newcommand{\www}{\mathrm{\pmb{w}}}
\newcommand{\mmm}{\mathrm{\pmb{m}}}
\newcommand{\rrr}{\mathrm{\pmb{r}}}
\newcommand{\CI}{\scriptscriptstyle{(\mathrm{CI}})}
\newcommand{\ML}{\scriptscriptstyle{(\mathrm{ML}})}
\newcommand{\EXP}{\mathbb{E}}
\newcommand{\PR}{\mathbb{P}}
\newcommand{\TR}{\mathrm{Tr}}
\newcommand{\VEC}{\mathrm{vec}}
\newcommand{\SUP}{\mathrm{sup}}
\newcommand{\INF}{\mathrm{inf}}
\newcommand{\DET}{\mathrm{det}}
\newcommand{\TT}{\mathrm{T}}
\newcommand{\C}{\mathbb{C}}
\newcommand{\RR}{\mathbb{R}}
\newcommand{\D}{\mathcal{D}}
\newcommand{\SPSK}{\mathrm{-S}\mathrm{PSK}}
\newcommand{\SQPSK}{\mathrm{-S}\mathrm{QPSK}}
\newcommand{\SAPSK}{\mathrm{-S}\mathrm{APSK}}
\newcommand{\SkPSK}{\mathrm{-S}^k\mathrm{PSK}}
\newcommand{\SkQPSK}{\mathrm{-S}^k\mathrm{QPSK}}
\newcommand{\SkAPSK}{\mathrm{-S}^k\mathrm{APSK}}
\newcommand{\PSNR}{\mathrm{PSNR}}
\newcommand{\SNR}{\mathrm{SNR}}
\newcommand{\LLR}{\mathrm{LLR}}
\newcommand{\diag}{\mathop{\mathrm{diag}}}

\newlength{\dhatheight}
\newcommand{\doublehat}[1]{%
	\settoheight{\dhatheight}{\ensuremath{\hat{#1}}}%
	\addtolength{\dhatheight}{-0.3ex}%
	\hat{\vphantom{\rule{1pt}{\dhatheight}}%
		\smash{\hat{#1}}}}

\maketitle

\begin{abstract}

In this paper, we propose a low-complexity method to approximately solve the SINR-constrained optimization problem of symbol-level precoding (SLP). First, assuming a generic modulation scheme, the precoding optimization problem is recast as a standard non-negative least squares (NNLS). Then, we improve an existing closed-form SLP (CF-SLP) scheme using the conditions for nearly perfect recovery of the optimal solution support, followed by solving a reduced system of linear equations. We show through simulation results that in comparison with the CF-SLP method, the improved approximate solution of this paper, referred to as ICF-SLP, significantly enhances the performance with a negligible increase in complexity. We also provide comparisons with a fast-converging iterative NNLS algorithm, where it is shown that the ICF-SLP method is comparable in performance to the iterative algorithm with a limited maximum number of iterations. Analytic discussions on the complexities of different methods are provided, verifying the computational efficiency of the proposed method. Our results further indicate that the ICF-SLP scheme performs quite close to the optimal SLP, particularly in the large system regime. 

\end{abstract}

\section{Introduction}

In wireless multiuser multi-input multi-output (MU-MIMO) broadcast channels, precoding techniques can be employed in order to mitigate the channel-induced multiuser interference (MUI) via spatially pre-processing the users' data stream prior to transmission. This pre-processing, in the optimal case, is shown to achieve the capacity of the MU-MIMO broadcast channel \cite{dpc}. Beyond simple linear precoding schemes, such as (regularized) zero-forcing (ZF) \cite{vec_per}, in a practical scenario the precoding design usually aims at optimizing a certain objective function subject to some given system/user requirements; this kind of design is often called objective-oriented precoding optimization \cite{tb_convex}. Within a wide variety of objective-oriented design criteria, two closely-related formulations are frequently addressed, namely, signal-to-interference-plus-noise ratio (SINR)-constrained power minimization \cite{tb_opt,tb_sinr,tb_conic}, and the max-min SINR with power constraints \cite{tb_conic,tb_sol_str}, where ``power'' may refer to either {\emph{total}} or {\emph{per-antenna}} transmit power.

From a different point of view, multiuser precoding schemes can be classified broadly into two groups, namely, block-level (conventional) and symbol-level techniques. The conventional precoding typically exploits the channel state information (CSI) to mitigate the MUI, regardless of the instantaneous users' data symbols; see e.g. \cite{tb_opt}. The precoder then may be redesigned according to the channel coherence time. On the other hand, symbol-level precoding (SLP) takes advantage of the readily-available data information (DI) by converting the instantaneous MUI into a constructive signal component, lying onto the so-called constructive interference (CI) regions \cite{slp_chr,slp_con}. The symbol-level design, therefore, requires to be specifically optimized for every instantaneous realization of the users' symbols. In delay-sensitive wireless applications, online precoding computation may suffer from high computational complexity of the symbol-level design. Rather, an offline computation also leads to an unfavorable computation cost for high-order modulation schemes even with moderate number of users \cite{slp_multi,slp_tsp}. Nonetheless, the considerable performance improvement offered by a symbol-level precoder is motivating to find a more practical solution with a reasonable complexity.

Recently, a promising effort has been made towards low complexity (sub-optimal) solutions for various types of the SLP design problem. The authors in \cite{slp_practical} propose an iterative method with a closed-form update equation for the max-min SINR SLP, where the algorithm is shown to converge to the optimal solution in a few iterations. In \cite{slp_cf}, a closed-form sub-optimal solution is obtained for the SINR-constrained power minimization SLP using the Karush-Kuhn-Tucker (KKT) optimality conditions. In another recent work \cite{slp_jev}, the SINR-constrained power minimization SLP is addressed with strict phase constraints on the received signals, and a low complexity approximate method is suggested for this particular case. However, the major drawback of the two latter methods is poor performance of the approximate solution for large numbers of transmit antennas and users.

In this paper, we revisit the SINR-constrained power minimization SLP problem assuming a generic modulation scheme with distance-preserving CI regions (DPCIR) (Section \ref{sec:sys}). The original formulation can be transformed into an equivalent non-negative least squares (NNLS) problem (Section \ref{sec:prob}). The NNLS representation enables us to derive a low-complexity approximate solution in a systematic way (Section \ref{sec:nnls}). This solution, which improves the method presented in \cite{slp_cf}, simply applies a validation step before calculating the final solution. Despite a slight increase in complexity, the new method shows noticeable performance gains. In particular, unlike \cite{slp_cf}, the gap to the optimal SLP remains almost steady with enlarging the system. It is further shown that the new method can be used as an alternative to (even fast-converging) NNLS algorithms, especially when complexity is a practical design limitation. 

\noindent{\bf{Notations:}} We use uppercase and lowercase bold-faced letters to denote matrices and vectors, respectively. The sets of real and complex numbers are represented by $\mathbb{R}$ and $\mathbb{C}$. For a matrix $\A$, $\mathcal{R}(\A)$ represents the column space of $\A$. $\mathrm{diag}(\cdot)$, or $\mathrm{blkdiag}(\cdot)$, represents a square (block) matrix having main-diagonal (block) entries and zero off-diagonals. For a set $\mathrm{S}$, $|\mathrm{S}|$ denotes the cardinality of $S$. Given two vectors $\x$ and $\yyy$ with equal dimensions, $\x\succeq\yyy$ (or $\x\succ\yyy$) denotes the entrywise inequality. $\|\cdot\|_2$ represent the vector Euclidean norm. $\pmb{I}$ and $\OOO$ respectively stand for the identity matrix and the zero matrix (or the zero vector, depending on the context) of appropriate dimensions. The operator $\otimes$ stands for the Kronecker product.

\section{System Model and CI Constraints}\label{sec:sys}

We consider an MU-MIMO broadcast channel in which a common transmitter (e.g., a base station), equipped with $N$ antennas, serves $K$ single-antenna users by sending independent data streams, where $K\leq N$. We denote by row vectors $\h_k\in\C^{1\times N}, k=1,...,K,$ the instantaneous (frequency-flat) fading channels of the transmit/receive antenna pairs. Focusing on a specific symbol instant, in the downlink transmission, independent data symbols $\{s_k\}_{k=1}^K$ are intended for different users, where the symbol $s_k$ corresponds to the $k$th user.

The set of desired symbols for all $K$ users needs to be mapped to $N$ transmit antennas, yielding the transmit signal $\uuu=[u_1,\ldots,u_N]^T\in\C^{N\times1}$. This mapping is done by means of a multiuser precoding module. In this paper, we adopt a symbol-level precoding (SLP) scheme. Thereby, the optimal transmit vector $\uuu$ is directly obtained as a result of an objective-oriented precoding optimization on a symbol-level basis. At the receiver of the $k$th user, the observed signal can be expressed as
\begin{equation}\label{eq:sys}
r_k = \h_k\uuu+z_k, \; k=1,...,K,
\end{equation}
where $z_k$ represents the additive circularly symmetric complex Gaussian noise distributed as $z_k\sim\mathcal{CN}(0,\sigma_k^2)$. The $k$-th user may use the maximum-likelihood (ML) single-user detector to optimally detect its desired symbol $s_k$; nevertheless, the structure of the receiver is independent of the precoder design. 
In the rest, we adopt the equivalent real-valued notations
$$\tilde{\uuu}\!=\!\begin{bmatrix}\REAL\{\uuu\} \\ \IMAG\{\uuu\}\end{bmatrix}, \HHH_k\!=\!\begin{bmatrix}
\REAL\{\h_k\} \; -\IMAG\{\h_k\}\\
\IMAG\{\h_k\} \quad\;\: \REAL\{\h_k\}
\end{bmatrix},\s_k\!=\!\begin{bmatrix}\REAL\{s_k\} \\ \IMAG\{s_k\}\end{bmatrix},$$
where $\tilde{\uuu}\in\mathbb{R}^{2N\times1}$, and $\HHH_k\in\mathbb{R}^{2\times2N}$ and $\s_k\in\mathbb{R}^{2\times1}$ for all $k=1,...,K$. Clearly, $\HHH_k \tilde{\uuu}=[\REAL\{\h_h\uuu\},\IMAG\{\h_h\uuu\}]^T$.

To exploit the DI in a symbol-level precoded broadcast, one needs to design the transmit signal such that each (noise-free) received signal $\HHH_k \tilde{\uuu}$ is observed within a pre-defined region corresponding to the intended symbol, called constructive interference region (CIR). The CIRs, which are modulation-specific regions, have been defined in several ways in the literature; see, e.g., \cite{slp_chr,slp_con,slp_gen}. As mentioned earlier, we focus on the so-called DPCIRs \cite{slp_gen}, which are presented in a generic form that is applicable to any given (two-dimensional) modulation scheme.

 For the sake of simplicity of analysis, and without loss of generality, we assume that identical modulation schemes are employed for all $K$ users. The associated symbol constellation is represented by $\mathrm{X}=\{\x_m:\x_m\in\mathbb{R}^{2\times1}\}_{m=1}^M$, where $\mathrm{X}$ is an equiprobable set with unit average power. We denote by $\mathrm{bd}(\mathrm{X})$ and $\mathrm{int}(\mathrm{X})$, respectively, the sets of boundary and interior points of $\mathrm{X}$. It has been shown in \cite{slp_tsp} that any $\x\in\mathbb{R}^{2\times1}$ belonging to the DPCIR of $\x_m$ satisfies
\begin{equation}\label{eq:dpcir}
\A_{m} \left(\x - \x_{m}\right) \succeq \OOO, \;\; \text{if} \;\, \x_m \in\mathrm{bd}(\mathrm{X}),\\
\end{equation}
or $\A_{m} \left(\x - \x_{m}\right) = \OOO$ otherwise, where $\A_{m} = [\aaa_{m,1},\aaa_{m,2}]^T = [\x_{m} - \x_{m,1},\x_{m} - \x_{m,2}]^T\in\mathbb{R}^{2\times2}$ contains the normal vectors of distance-preserving boundaries,
with $\x_{m,1}$ and $\x_{m,2}$ denoting two (specific) neighboring constellation points of $\x_{i}$. Let $\Deee_m\in\RR_+^{2\times 1}$ be a non-negative vector, then the representation in \eqref{eq:dpcir} can be equally expressed by
\begin{equation}\label{eq:dpcir2}
\A_{m} \left(\x - \x_{m}\right) = \Deee_m,\;\text{where}\,
\begin{cases}
\Deee_m\succeq\OOO, & \; \x_m\in\mathrm{bd}(\mathrm{X}),\\
\Deee_m=\OOO, & \; \x_m\in\mathrm{int}(\mathrm{X}),
\end{cases}
\end{equation}
For a detailed discussion on the characteristics of DPCIRs, the interested readers are kindly referred to \cite{slp_tsp}.



\section{SINR-Constrained Power Minimization SLP}\label{sec:prob}

In this section, we overview the instantaneous (per-symbol) power minimization problem constrained by CIRs and given SINR requirements $\gamma_k,k=1,...,K$. The users' intended symbols $\{\s_k\}_{k=1}^K$ are taken from the set of points $\{\x_m\}_{m=1}^M$ in $\mathrm{X}$. We denote by $m_k$ the index of the constellation point that corresponds to $\s_k$, i.e., $\s_k=\x_{m_k}$ where $\x_{m_k}\in \mathrm{X}$ and $m_k\in\{1,...,M\}$. By assuming DPCIRs, the convex representation in \eqref{eq:dpcir2} can be used to imply the CI constraint in the optimization problem. By substituting $\HHH_k \tilde{\uuu}$ for $\x$ and replacing the scaled symbol $\sigma_k\sqrt{\gamma_k}\;\x_{m_k}$, the joint CI/SINR constraint for the $k$-th user is expressed by
\begin{equation}\label{eq:ci}
\A_{m_k} \left(\HHH_k \tilde{\uuu} - \sigma_k\sqrt{\gamma_k}\,\x_{m_k}\right) = \Deee_{m_k},\; \Deee_{m_k}\succeq\OOO,
\end{equation}
where $\Deee_{m_k}=\OOO$ is imposed for $\x_{m_k}\in\mathrm{int}(\mathrm{X})$.
Let $\WWW$ be a square binary weighting matrix defined as
\begin{equation}\label{eq:w}
\WWW\triangleq\diag(w_{m_1},...,w_{m_K})\otimes\I_2, \;\; w_{m_k} = \begin{cases}
1, & \; \x_m\in\mathrm{bd}(\mathrm{X}),\\
0, & \; \x_m\in\mathrm{int}(\mathrm{X}).
\end{cases}
\end{equation}
By stacking the constraints in \eqref{eq:ci} for all $k\in\{1,...,K\}$ into a compact matrix form, we have
\begin{equation}\label{eq:lmeall}
\A (\tilde\HHH \tilde{\uuu} - \Sigmaaa\Gammaaa^{1/2}\,\tilde{\x}) = \WWW\Deee, \; \Deee\succeq\OOO,
\end{equation}
where $\tilde{\HHH} \triangleq [\HHH_1^T,...,\HHH_K^T]^T$, $\A \triangleq \mathrm{blkdiag}(\A_{m_1},...,\A_{m_K})$, $\Sigmaaa \triangleq \diag(\sigma_1,...,\sigma_K)\otimes \I_2$, $\Gammaaa \triangleq \diag(\gamma_1,...,\gamma_K)^T\otimes \I_2$, $\tilde{\x}\triangleq[\x_{m_1},...,\x_{m_K}]^T$, $\Deee\triangleq[\Deee_{m_1},...,\Deee_{m_K}]^T$, and $(\cdot)^{1/2}$ denotes the matrix square root. Then, the optimal symbol-level precoded transmit vector can be obtained by the following lemma \cite{slp_tsp}.

\begin{lemma}\label{lem:0}
The minimum-norm vector satisfying the DPCIR constraint of \eqref{eq:lmeall} is given by
\begin{equation}\label{eq:u}
\tilde{\uuu}^* = \tilde{\HHH}^\dagger\left(\Sigmaaa\Gammaaa^{1/2} \tilde{\x} + \A^{-1}\WWW\Deee^*\right),
\end{equation}
where $\Deee^*$ is the optimal solution to the following non-negative least squares (NNLS) problem
\begin{equation}\label{eq:pm4}
\underset{\Deee\succeq\OOO}{\min} \quad \|\tilde{\HHH}^\dagger\Sigmaaa\Gammaaa^{1/2} \tilde{\x} + \tilde{\HHH}^\dagger \A^{-1}\WWW\Deee\|^2.
\end{equation}
\end{lemma}

It follows from Lemma \ref{lem:0} that the design problem of interest can be tackled through solving the NNLS optimization in \eqref{eq:nnls}. Furthermore, denoting $\BBB\triangleq-\tilde{\HHH}^\dagger \A^{-1} \WWW$ and $\yyy\triangleq\tilde{\HHH}^\dagger\Sigmaaa\Gammaaa^{1/2} \tilde{\x}$, the NNLS problem \eqref{eq:pm4} can be written in the standard form as
\begin{equation}\label{eq:nnls}
\underset{\Deee\succeq\OOO}{\min} \quad \|\yyy-\BBB\Deee\|^2,
\end{equation}
 The NNLS problem, unlike its unconstrained counterpart, does not in general admit a closed-form solution due to the non-negativity constraints. Various efficient algorithms to solve an NNLS can be found in the literature on iterative optimization, such as the well-known active set based method proposed by Lawson and Hanson \cite{nnls_lawson}, the fast NNLS algorithm (FNNLS) \cite{nnls_fast}, and those based on projected/proximal gradient method \cite{nnls_gradient, nnls_accelerated,nnls_proximal}. However, an NNLS algorithm, in the best known case, requires tens of iterations to converge. For instance, the accelerated gradient method have a linear convergence rate of $\mathcal{O}(n^{-2})$, where $n$ is the number of iterations. With a convex objective function, this translates to a worst-case complexity bound of $\mathcal{O}(\epsilon^{-1/2})$ to reach an $\epsilon$-optimal solution. As an illustrative example, using the accelerated projected gradient descent, it takes nearly $100$ iterations to have a residual of $10^{-3}$ with respect to the optimum. In a symbol-level design application, this process needs to be done as repeatedly as either the frame length or the total number of possible symbol realizations for $K$ users, i.e., $M^K$. This motivates us to still be looking for a more computationally efficient, though possibly approximate, solution for the SLP design problem.

\section{Low-Complexity Solution for NNLS-based SLP}\label{sec:nnls}

The main goal of this section is to obtain a low-complexity solution for the NNLS design formulation of SLP in \eqref{eq:nnls}. We first proceed by reviewing some basic mathematical analysis on the NNLS problem.

Let $\Deee^*=[\delta^*_1,...,\delta^*_{2K}]^T$ denote the minimizer corresponding to the standard NNLS formulation in \eqref{eq:nnls}. We refer to the set of indices $j$ for which $\delta^*_j>0$ as the support of $\Deee^*$, or the optimal support, denoted by $\Lambda^*=\{j:\delta^*_j>0\}$. Given the optimal support $\Lambda^*$, the minimizer of \eqref{eq:nnls} can be simply computed by $({\BBB_{\Lambda^*}})^\dagger\yyy$ with appropriate zero-padding, where $\BBB_{\Lambda^*}$ denotes the matrix composed of those columns of $\BBB$ associated with the indices in $\Lambda^*$. In other words, finding $\Lambda^*$ is as complex as solving \eqref{eq:nnls} for the optimal solution. Therefore, one may attempt to solve \eqref{eq:nnls} equivalently by perfectly identifying $\Lambda^*$. This is in fact the underlying idea of active set methods, where at each iteration some constraints are set to be active (i.e., zero-valued in our context), while the other constraints are used in the update equation. However, here we are interested in having an estimate of $\Lambda^*$, say $\hat{\Lambda}$, obtained in a non-iterative manner. This allows us to derive an approximate solution $\hat{\Deee}$ in an explicit form. Our proposed method is essentially based on the following lemma which states sufficient conditions for nearly perfect recovery of the optimal support \cite{nnls_perfect}.
\begin{lemma}\label{lem:1}
	Let $\Lambda$ be a subset of column indices of the matrix $\BBB$ with $|\Lambda|\leq 2K$, and the columns associated with the indices in $\Lambda$ are linearly independent. Let $\Deee^*\succeq\OOO$ be the minimizer of $\|\yyy-\BBB\Deee\|^2$. Then, $\Lambda$ coincides with the support of $\Deee^*$ if
	$$\mathrm{C}1:\enspace {\BBB_\Lambda}^\dagger\yyy\succ\OOO, \quad \text{and} \quad \mathrm{C}2:\enspace \yyy^T\PPP^\perp_\Lambda\bbb_i<\OOO,\;\forall i\in\Lambda^\mathrm{c},$$
	where $\PPP^\perp_\Lambda$ is the projector onto the orthogonal complement of $\mathcal{R}(\BBB_\Lambda)$, $\bbb_i$ denotes the $i$th column of $\BBB_\Lambda$, and $\Lambda^\mathrm{c}=\{1,...,2K\}-\Lambda$.
\end{lemma}
Based on Lemma \ref{lem:1}, both the conditions $\mathrm{C}1$ and $\mathrm{C}2$ together are sufficient for a candidate support $\Lambda$ to be optimal. In fact, $\mathrm{C}1$ measures if the resultant solution satisfies the positivity constraint (notice that the constraint cannot be satisfied with equality due to the definition of support), while the projection in $\mathrm{C}2$ can be read as the deviation of $\yyy$ from the column space of $\BBB_\Lambda$. In other words, $\mathrm{C}1$ is required to validate the columns already indexed in $\Lambda$, whereas $\mathrm{C}2$ assesses the possibility of including any of the columns belonging to $\Lambda^\mathrm{c}$. Armed with these two conditions, we are ready to approximately solve the NNLS problem in \eqref{eq:nnls}, as will be explained in the sequel.

\subsection{The Proposed Approximate Solution}

First, we exploit the projection condition $\mathrm{C}2$ to produce a rough estimate of $\Lambda^*$. Let
$$d_i \triangleq \yyy^T\PPP^\perp_\Lambda\bbb_i, \; i=1,...,2K,$$
where $$\PPP^\perp_\Lambda=\I-\BBB_{\Lambda}\left({\BBB_\Lambda}^T\BBB_{\Lambda}\right)^{-1}{\BBB_\Lambda}^T.$$
Treating the entire columns of $\BBB$ as candidate columns to be indexed in $\Lambda$, we assume $\Lambda^\mathrm{c}=\{1,...,2K\}$, yielding $\PPP^\perp_{\Lambda}=\I$. Hence,
\begin{equation}\label{eq:step10}
d_i=\yyy^T\bbb_i, \; i=1,...,2K.
\end{equation}
With the inner products in \eqref{eq:step10}, we define $\hat{\Lambda} \triangleq \left\{i:d_i > 0\right\}$ with $|\hat{\Lambda}|=L_1$, which builds our initial estimate of $\Lambda^*$. Notice that the conditions in \eqref{eq:step10} are similarly implied from the KKT optimality conditions, as discussed in \cite{slp_cf}. Next, we validate this estimate by excluding those columns in $\hat{\Lambda}$ that result in negative elements for $\Deee$, i.e.,
\begin{equation}\label{eq:step12}
\doublehat{\Lambda} \triangleq \left\{l:l\in\hat{\Lambda},\left[(\BBB_{\hat{\Lambda}})^\dagger\yyy\right]_l>0\right\},
\end{equation}
where $[\cdot]_l$ denotes the $l$th element of an input vector. Clearly, we have $|\doublehat{\Lambda}|\triangleq L_2\leq L_1$, which reduces the possibility of having negative elements in the final solution as a result of the additional validation step in \eqref{eq:step12}. Our simulations indicate that in most cases $\doublehat{\Lambda}$ gives a more accurate estimate of the optimal support $\Lambda^*$, compared to that given by $\hat{\Lambda}$, as we will see in the next section. Notice, however, that the positivity constraints may still be violated even after the validation step in \eqref{eq:step12} since the remaining set of columns in $\doublehat{\Lambda}$ does not necessarily guarantee that  $(\BBB_{\hat{\vphantom{\rule{1pt}{5.8pt}}\smash{\hat{\Lambda}}}})^\dagger \yyy\succ\OOO$. Therefore, one needs to ignore all the negative elements in the final solution, if any. More precisely, due to the fact that $\mathcal{R}\big(\BBB_{\hat{\vphantom{\rule{1pt}{5.8pt}}\smash{\hat{\Lambda}}}}\big)\subseteq\mathcal{R}\big(\BBB_{\hat{\Lambda}}\big)$, perfect recovery of the optimal support is possible only if $\Lambda^*\subseteq\hat{\Lambda}$. In such case, we obtain $(\BBB_{\hat{\vphantom{\rule{1pt}{5.8pt}}\smash{\hat{\Lambda}}}})^\dagger \yyy\succ\OOO$ and $\doublehat{\Lambda}$ is the optimal support. Consequently, the approximate solution $\hat{\Deee}=[\hat{\delta}_1,...,\hat{\delta}_{2K}]^T$ can be represented as a zero-padded version of the vector $(\BBB_{\hat{\vphantom{\rule{1pt}{5.8pt}}\smash{\hat{\Lambda}}}})^\dagger \yyy$, i.e.,
\begin{equation}\label{eq:zp}
\hat{\delta}_l =\max\left\{\left[(\BBB_{\hat{\vphantom{\rule{1pt}{5.8pt}}\smash{\hat{\Lambda}}}})^\dagger \yyy\right]_l,0\right\}, \; l\in\doublehat{\Lambda},
\end{equation}
and $\hat{\delta}_l=0$ otherwise, for $l\!=\!1,...,2K$. Having an explicit expression for $\hat{\Deee}$, the corresponding transmit vector is readily computable by replacing $\hat{\Deee}$ in \eqref{eq:u}.

\subsection{Computational Complexity Analysis}

\begin{algorithm}[t]
	\caption{\small{APGD algorithm for the SLP NNLS problem \eqref{eq:nnls}}}
	\label{alg:1}
	
	\begin{algorithmic}[1]
		\linespread{1}
		\State \bf{input} $\BBB, \yyy, n_\mathrm{max}$
		\State \bf{initialize} $\Varrr_0 \!=\! \Deee_0 \!\in\!\mathbb{R}_+^{2K\times1}, \QQQ\!=\!\I - \frac{\BBB^T\BBB}{\|\BBB^T\BBB\|_F}, \Phiii\!=\!\frac{\BBB^T\yyy}{\|\BBB^T\BBB\|_F}$
		\State set $\eta=\frac{1-\sqrt{\kappa}}{1+\sqrt{\kappa}}, \kappa = \frac{\sigma_\mathrm{max}(\BBB)}{\sigma_\mathrm{min}(\BBB)},n=0$
		\While {$n < n_\mathrm{max}$}
		\State $n = n + 1$
		\State $\Deee_n = \max\left\{\QQQ\Varrr_{n-1} + \Phiii,\OOO\right\}$
		\State $\Varrr_n = \Deee_n + \eta (\Deee_n-\Deee_{n-1})$
		\EndWhile
	\end{algorithmic}
	{\footnotesize{note: $\sigma_\mathrm{max}(\cdot)$ and $\sigma_\mathrm{min}(\cdot)$ respectively denote the maximum and minimum singular values of an input matrix}}.
\end{algorithm}

We compare the computational complexity of the proposed method with an iterative NNLS algorithm. As our benchmark for comparison, we consider the accelerated projected gradient descent (APGD) algorithm \cite{nnls_gradient}. The pseudocode of APGD to (approximately) solve the NNLS \eqref{eq:nnls} via a limited number of iterations is given in Algorithm \ref{alg:1}. We evaluate the worst-case complexity in terms of the number of arithmetic operations. For an iterative method, this can be interpreted as the required number of operations until the stopping condition is met. 

The main loop of APGD is preceded by an initialization step performing two matrix multiplications and one singular value decomposition (SVD) with complexity orders of $K^2N$, $KN$ and $K^3$, respectively. Within the main loop, the per-iteration complexity is dominated by a matrix multiplication of order $K^2$. To be more accurate, the complexity of APGD depends also on the convergence specifications, e.g., the condition number of $\BBB$; however, we consider only those complexity terms directly relating to the problem size. On the other hand, the dominant arithmetic operations in \eqref{eq:step10}, \eqref{eq:step12} and \eqref{eq:zp}, i.e., $2K$ vector multiplications and two matrix pseudo-inversions, result in computation costs of order $KN$ and $N(L_1^2+L_2^2)$, respectively, for the proposed method. Remark that the closed-form solution in \cite{slp_cf} can be implemented in an equivalent way using \eqref{eq:step10} and \eqref{eq:zp}; therefore we assess the complexity of \cite{slp_cf} based on the method of this paper.

\begin{table}[hbt!]
	\caption{Complexities of different NNLS-based SLP solutions.}
	\label{tab:co}
	\centering
	\renewcommand{\arraystretch}{1.3}
	\begin{tabular}{ll}
		\hline
		{\bf Solution Method} & {\bf Complexity Order}\\
		\hline
		
		APGD algorithm \cite{nnls_gradient} & $K^2.\,\mathcal{O}\left(K+N\right)+\mathcal{O}\left(K^2\right)\epsilon^{-1/2}$
		\\
		\hline
		Closed-form SLP \cite{slp_cf} & $N.\,\mathcal{O}\left(K+L_2^2\right)$
		\\
		\hline
		Improved closed-form SLP & $N.\,\mathcal{O}\left(K+L_1^2+L_2^2\right)$
		\\
		\hline
	\end{tabular}
\end{table}

In Table \ref{tab:co}, we summarize the dominating complexity orders of different methods, where that of the APGD corresponds to an $\epsilon$-optimal solution. Due to the sparsity-promoting nature of the NNLS problem \cite{nnls_sparsity}, in practice we have $L_2\leq L_1 \ll 2K$. Based on this observation and the results in Table \ref{tab:co}, we conclude that both closed-form methods (potentially) decrease the computation cost of the precoding design. In fact, even the complexity of the initialization step in APGD (without any iterations) is higher than the two closed-form methods.

\section{Simulation Results}\label{sec:sim}

In this section, we provide some simulation results to evaluate and compare the performances of various approaches to solve the SINR-constrained SLP problem. Our simulation setup is as follows. We consider a downlink MU-MIMO system with $N$ transmit antennas and $K$ (single-antenna) users, where $N/K\triangleq \beta$. For all users $k\in\{1,...,K\}$, we assume unit noise variances $\sigma_k^2=1$ and equal target SINRs $\gamma_k\triangleq\gamma$. The users' channel vectors $\{\h_k\}_{k=1}^K$ are independently generated following the standard circularly symmetric complex Gaussian distribution, i.e., $\h_k\sim\mathcal{CN}(\OOO,\I)$. The maximum number of iterations, $n_\mathrm{max}$, for the APGD algorithm is set to be $25$. The results are all averaged over $10^3$ channel coherence blocks each of length $10^3$ symbols. Later on in this section, we refer to the SLP methods of interest as:

\begin{itemize}
	\item[-] {\emph{ZF-SLP: symbol-level ZF, assuming $\Deee=\OOO$ in \eqref{eq:nnls}}}.
	\item[-] {\emph{Optimal SLP: optimal solution to \eqref{eq:nnls}}}.
	\item[-] {\emph{NNLS-SLP (APGD): solving \eqref{eq:nnls} via APGD algorithm}}.
	\item[-] {\emph{CF-SLP: closed-form approximate SLP solution of \cite{slp_cf}}}.
	\item[-] {\emph{ICF-SLP: Improved CF-SLP proposed in this paper}}.
\end{itemize}

\begin{figure}
	\centering
    \includegraphics[trim={0 0 0 .39in},clip,width=.55\columnwidth]{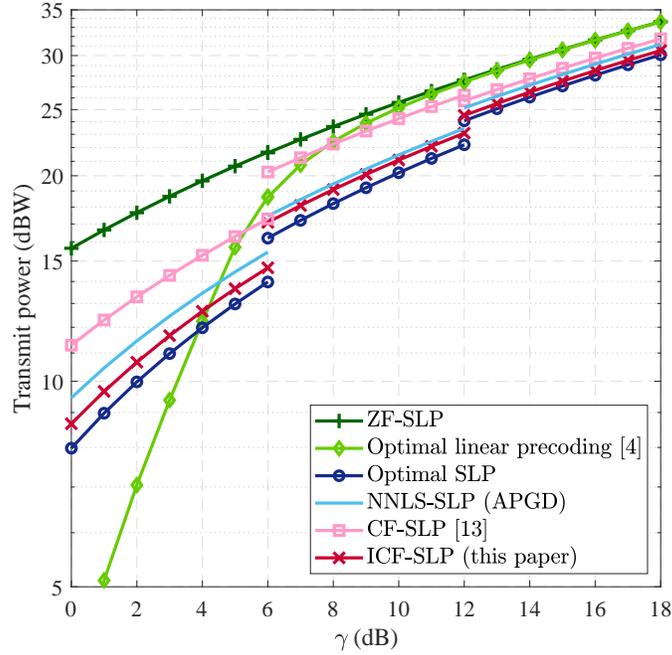}
	\caption{Transmit power versus target SINR with $N=K=8$. Three different modulations, namely QPSK, 8PSK and 16QAM, are respectively used in 0-6, 6-12 and 12-18 dB SINR ranges.}
	\label{fig:power_snr}
\end{figure}

In Fig. \ref{fig:power_snr}, the total transmit powers obtained from various precoding schemes are plotted as a function of target SINR, where three different modulations are assessed in the whole range of depicted SINR. The results correspond to a fully-loaded system with $N=K$. It can be seen that the ICF-SLP method improves the accuracy of the approximate solution by up to $3$ dB, compared to its naive counterpart CF-SLP. Further, ICF-SLP outperforms the NNLS-SLP method via APGD with $n_\mathrm{max}=25$. The observations show that both the methods have nearly the same complexity in this range of $K$. The promising fact about Fig. \ref{fig:power_snr} is that ICF-SLP performs well close to the optimal SLP with a far less complexity, as we will see next.

\begin{figure}
	\centering
	\includegraphics[width=.55\columnwidth]{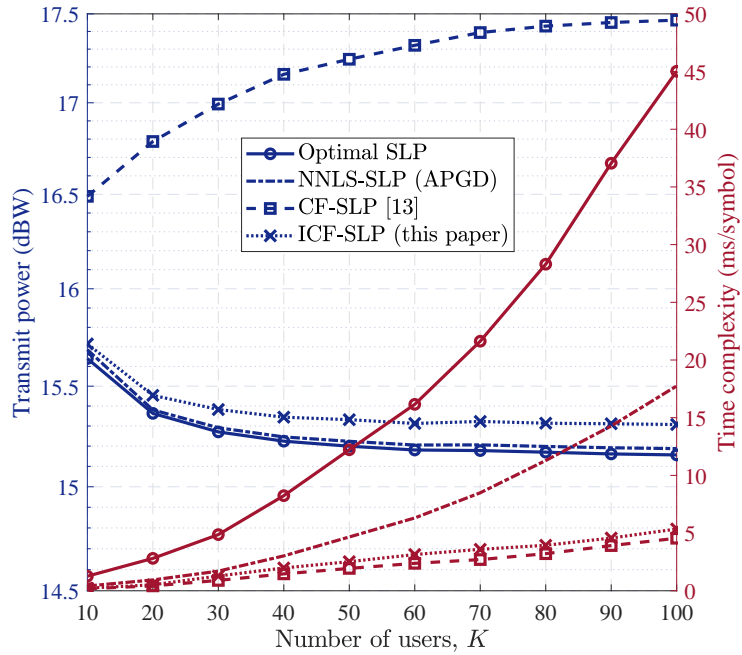}
	\caption{Transmit power and time complexity versus number of users, $\beta=6/5$. Same line types and markers as those in the legend refer to the right axes but with a different color.}
	\label{fig:power_time_nk}
\end{figure}

In another set of simulations for an under-loaded system with $\beta=6/5$, we evaluate the performance/complexity of different solution approaches to the SLP problem \eqref{eq:nnls}. The results are presented in Fig. \ref{fig:power_time_nk} as a function of the number of users $K$. The optimal SLP solution is obtained using the {\emph{lsqnonneg}} function of MATLAB, which uses the Lawson and Hanson active set method to solve NNLS. As it can be seen, the resulting performance of CF-SLP noticeably degrades with increasing $K$, whereas the proposed ICF-SLP shows a decreasing trend in transmit power (as that of the optimal SLP) for large system dimensions. The optimality gap of ICF-SLP with $K=100$ is just $0.15$ dBW. This improvement becomes of great significance when we consider also the time complexities of the two solutions; see Table \ref{tab:co}. Therefore, the time complexity results in Fig. \ref{fig:power_time_nk} are consistent with the analytical evaluations reported in Table \ref{tab:co}. This can be further verified through comparing ICF-SLP and the APGD-based NNLS-SLP method. The latter method has a higher complexity growth rate, which is theoretically proportional to $\mathcal{O}(K^2N)$ in the limiting case. This might suggest a performance-complexity tradeoff. However, notice that with $\eta_\mathrm{max}=25$, the dominating complexity order of the APGD algorithm in the large system limit stems from the initialization step, which is higher than the entire computation cost of ICF-SLP.

\section{Conclusion}\label{sec:con}

We proposed a low-complexity method to approximately solve the SLP power minimization problem with SINR constraints. Due to the required per-symbol computation, solving the SLP optimization problem for the exact solution may lead to an impractical transmitter complexity. To address this issue, we exploited the structure of an equivalent NNLS formulation of the original problem. We modified an existing approximate solution by applying a computationally efficient validation step before calculating the final solution. Based on our simulation results, this modification considerably reduces the loss with respect to the optimal solution, particularly in the large system regime. Further, the new approximate solution is shown to be comparable with the solution obtained from an iterative NNLS algorithm, from both performance and complexity points of view. It is, however, concluded that as far as a low-complexity implementation of SLP with a close-to-optimal performance is of concern, the proposed method provides a more efficient solution.


\end{document}